\documentclass[12pt]{article} 

\usepackage{amssymb,verbatim,graphicx,natbib}
\newcommand{\blank}{$\mbox{}$}
\newcommand{\bNotBmka}{\blank}
\newcommand{\eNotBmka}{\blank}
\usepackage{tikz}
\usetikzlibrary{trees}
\usepackage{algorithm}
\newcommand{\epsdir}{.}
\newtheorem{thm}{Theorem}[subsection]
\newtheorem{coroll}{Corollary}[subsection]
\newtheorem{lmm}{Lemma}[section]
\newcommand{\jpar}[1]{\par{\em #1}}
\newcommand{\qed}{{\hfill \framebox{$\mbox{}$} \par}}
\newcommand{\f}{\,f}
\newcommand{\p}{{\,{p}}}
\newcommand{\I}{I}
\newcommand{\E}{\,E}    
\renewcommand{\P}{{\mathsf P}}
\renewcommand{\aa}{A}
\newcommand{\bb}{B}
\newcommand{\cc}{C}
\newcommand{\dd}{D}
\newcommand{\pp}{P}
\newcommand{\calP}{\mathcal{P}} 
\newcommand{\ind}{\bot\hspace*{-6pt}\bot}  
\newcommand{\indep}{\ind}  
\newcommand{\exc}{\backslash}
\newcommand{\qbox}[1]{\quad\mbox{ #1 }\quad}
\newcommand{\jand}{\qbox{and}}
\newcommand{\jfor}{\qbox{for}}
\newcommand{\half}{\frac{1}{2}}
\renewcommand{\jpar}[1]{\par}
\newenvironment{keywords}{Keywords:}{\par}
\newenvironment{proof}{}{\qed}
\newcounter{examplecounter}
\newenvironment{example}{
 \refstepcounter{examplecounter}%
 {\par\noindent{\it Example} \arabic{examplecounter}.}
}{\par}

\begin{document}
\setlength{\baselineskip}{14pt}    
\title{ Synergy, suppression and immorality: forward differences of
  the entropy function }
\author{\small
\small Joe Whittaker${ ^1}$, Florian Martin${ ^2}$ and Yang Xiang${ ^2}$ \\ 
\small ${ ^1}$ Mathematics and Statistics, Lancaster University, UK, joe.whittaker@lancaster.ac.uk, \\
\small ${ ^2}$  Biological Sciences, PMI Research and Development, Neuchatel, Switzerland. 
}
\date{}
\maketitle{}
\begin{abstract}
Conditional mutual information  is important in the selection and
interpretation of graphical models.
Its empirical version is well known as a generalised likelihood ratio
test and that it may be represented as  a difference in entropy.
We consider the forward difference expansion of the entropy function
defined on all subsets of the variables under study.
The elements of this expansion are invariant to permutation of their
suffices and relate higher order mutual informations to lower order ones.
The third order difference is expressible as an, apparently
assymmetric, difference between a marginal and a conditional mutual
information.
Its role in the decomposition for explained information provides a
technical definition for synergy between three random variables.
Positive values occur when two variables provide alternative
explanations for a third; negative values, termed synergies, occur
when the sum of explained information is greater than the sum of its
parts.
Synergies tend to be infrequent; they connect the seemingly unrelated
concepts of suppressor variables in regression, on the one hand, and
unshielded colliders in Bayes networks (immoralities), on the
other.
We give novel characterizations of these phenomena  that 
generalise to categorical variables and to higher dimensions.
We propose an algorithm for systematically computing low order
differences from a given graph.
Examples from small scale real-life studies indicate the potential of
these techniques for empirical statistical analysis.
\end{abstract}
\par
\begin{keywords}
Bayes network;
Conditional mutual information;
Imsets;
Mobius inversion;
Suppressor variable;
Unshielded collider.
\end{keywords}
\par
\section{\sffamily\large Introduction}
The independence of two random variables is denoted by 
${ X_1\indep{}X_2}$, \cite{dawid1979conditional}, and the conditional
independence of these two, given a third, by ${ X_1\indep{}X_2\mid{}X_3}$.
The marginal mutual information  of two random variables and the
conditional mutual information  of two variables given a third
are, in terms of the joint probability density or mass function,
{\begin{eqnarray}\label{eq:cmi}
\I_{ 12} = \inf(X_{1}\indep{}X_{2}) = \E \log \frac{f_{ 12}}{f_{ 1}f_{ 2}}
\;\mbox{ and }\;
 \I_{ 12\mid{}3} = \inf(X_{1}\indep{}X_{2}\mid{}X_{3}) 
=
\E \log \frac{f_{ 12|3}}{f_{ 1|3}f_{ 2|3}}{}.
\end{eqnarray}}\blank\newline
\par{}
The difference in these measures is
{\begin{eqnarray}\label{eq:delta}
\inf(X_{1}\indep{}X_{2}) -
\inf(X_{1}\indep{}X_{2}\mid{}X_{3})
&=& 
-\E \log \frac{f_{ 123}f_{ 1}f_{ 2}f_{ 3}}{ f_{ 12}f_{ 13}f_{ 23}}.
\end{eqnarray}}\blank\newline
The key property is that the right hand side  is symmetric to any
permutation of suffices ${ 1,2,3}$ even though the left does not appear to
be.
Define ${ \delta_{ 123}}$ by the right hand side expression.
\par{}
Conditional independence and mutual information lie at the foundations
of graphical models, for texts see
\cite{koller2009probabilistic,lauritzen1996graphical,whittaker1990graphical}.
The seminal citation for the separation properties of undirected
graphical models is \cite{darroch1980markov}.
\cite{pearl1988probabilistic} made the big step in establishing
acyclic directed graphs and the concept of d-separation for Bayes
networks.
In the class of these directed graphs certain subsets are
probabilistically (Markov) equivalent, and an important theorem is
that the skeleton and its unshielded colliders (or immoralities)
specify the equivalence classes.
The criterion for this and the generalisation to chain graphs was independently established by \cite{verma1990equivalence} and
\cite{frydenberg1990chain}, and later developed by
\cite{andersson1997characterization}.
\par{}
Suppressor effects in multiple regression were first elucidated by
\cite{horst1941role}.
The phenomenon arises when the dependent variable has a smaller
prediction error by including an additional explanatory variable that
has no (or little) explanatory effect when used by itself; often this
is manifest in enhanced regression coefficients.
There is a  social science literature concentrated in educational
and psychological testing theory that has an interest in suppression
because of its concern to design experiments that make predictions
more precise.
\cite{ludlow2014suppressor} gives a substantial review of this area.
and so we just mention a few other references:
\cite{mcnemar1945mode},
\cite{voyer1996relation},
\cite{maassen2001suppressor},
\cite{shieh2006suppression}.
The well known structural equations text \cite{kline2011principles}
cites suppression among one of the fundamental concepts of regression.
\par
The technical literature on alternative ways
to define and explain suppression includes
\cite{velicer1978suppressor},
\cite{bertrand1988quirk},
\cite{smith1992suppressor},
\cite{mackinnon2000equivalence}, 
\cite{shieh2001inequality}.
More recently \cite{friedman2005graphical}, give a survey and point
out that the term synergism follows a suggestion of
\cite{hamilton1988letter}.
This literature distinguishes several types of  suppression.
Classical suppression: ${ X_2}$, say, is the suppressor variable, it is
uncorrelated (or nearly so) with ${ Y}$, but adds to the predictive power
of the regression ${ Y}$ on ${ X_1}$ when both included.
Negative suppression: both ${ X_1}$, ${ X_2}$ have a positive zero-order
correlations with ${ Y}$, and correlate positively with each other, but
one has a negative coefficient in the regression of ${ Y}$ on both.
Reciprocal suppression: both variables are marginally good predictors
of ${ Y}$, but are negatively correlated.
\par There are several seemingly different indicators of suppression,
which are varyingly described by conditions on the correlations
(marginal, multiple, partial), or in terms of regression and
correlation coefficients, or in terms of explained variance, or even
in terms of a rather confusing semi-partial correlation introduced by
Velicer.
All authors give conditions for three variable regression scenario, some
attempt to generalise to ${ p}$-variables, and some explanations are
geometric; most however reduce to conditions on correlation
coefficients.
That suppression is usually presented as a three dimensional
correlation phenomenon does not make clear how to measure its
strength, or how to generalise to higher dimensions; or how to
generalise to other distributions.
\par{}
Our contribution is to show the 3rd-order forward difference of
(\ref{eq:delta}) relates the seemingly unrelated topics of immorality
and suppression in a natural way.
The condition for suppression is that ${ \delta_{ 123}<0}$;  noting the
phrase `the whole regression can be greater than the sum of its parts'
in the title of \cite{bertrand1988quirk} suggests that synergy is a
good synonym for the triple ${ 123}$.
The condition for an unshielded collider (immorality) at ${ 3}$ is that
${ \delta_{ 123}<0}$ and ${ \delta_{ 12}=0}$.
\par
To set this within a wider framework we write down forward difference
expansion for the entropy function, and use Mobius inversion to
calculate the differences given the entropies.
All forward differences are invariant to permutation of their
suffices.
Marginal mutual informations are second order differences and
conditional measures have additive expressions in terms of the second
and higher order forward differences.
Higher order differences, are made more tractable by defining
conditional forward differences.
\par
We interpret the negative third order forward differences 
as synergies.
Classic examples of graphical models in low dimensions illustrate the
role of forward differences in interpretation of the model.
A computing scheme for 3rd-order elements from a given graph based on
cluster nodes is used to investigate empirical data for synergies.
\par
The  forward differences of the entropy provides  a wider framework
to explore suppression and immorality.
This setting explains why the essence of both phenomena concerns
exactly three variables;
and why suppression is symmetric.
It distinguishes suppression from both mediation and confounding where
${ \delta}$ is positive.
It generalises the notion of suppressor variables to higher dimensions
and to other distributions, for instance to categorical data.
It gives an alternative characterization of immoralities (unshielded
colliders).
\bNotBmka 
\par{\em\sf Plan of the paper:}
In Section 2 we define the forward difference expansion of the entropy
function and elaborate its properties.
In Section 3 we make the connection to suppressor variables in
regression and immoralities in Bayes networks, and give alternative
characterizations of these phenomena.
In Section 4 we consider more detailed applications to the categorical
and continuous data and examples from small scale empirical studies.
Proofs are collected in the Appendix.
An algorithm for systematically computing low order differences from a
given graph is provided in Supplementary Material.
\eNotBmka 
\par
\section{\sffamily\large Forward differences of the entropy}\label{sect:fwddiff}
\par
\subsection{\sffamily Preliminaries}
The  nodes in ${ \pp=\{1,2,\dots,p\}}$ correspond to random variables
${ X_1,X_2,\dots,X_p}$ having a joint distribution.
For subsets ${ \aa,\bb,\cc}$ of the power set
${ \calP=\{\phi,\{1\},\{2\},\dots,\pp\}}$ conditional independence 
statements of the form ${ X_{\aa}\indep{}X_{\bb}\mid{}X_{\cc}}$ where
${ X_{\aa}}$ refers to the vector ${ (X_{i}; i\in\aa)}$ simplify the
dependency structure of ${ X(\equiv{}X_{\pp})}$.
\par
The entropy function, ${ h}$, is defined on ${ \calP}$ by
${ h_{\aa}=-\E\log\f_{ \aa}(X_{\aa})}$, 
where ${ \f}$ is the derivative of the joint probability measure.
Without loss of generality we assume this is always well defined
for, if not, we may replace it by the (negative) relative entropy
${ -\E\log\f_{ \aa}(X_{\aa})/}$ ${ \prod_{ i\in{}\aa}\f_{ i}(X_{i})}$,
termed the multi-information by  \cite{studeny2005probabilistic}.
For a ${ p}$-dimensional mass point distribution the entropy is 
${ h_{\aa}=-\sum_{x_{\aa}}\p_{ \aa}(x_{\aa})\log\p_{ \aa}(x_{\aa})}$
where ${ \p_{ \aa}}$ is the   mass function on the margin
determined by ${ \aa}$.
This is always non-negative.
For a ${ p}$-dimensional multivariate Normal distribution with mean zero
and correlation matrix ${ \Sigma}$ the entropy is
${ h_{\aa}=1/2\log\det(\Sigma_{ \aa\aa})}$.
Any additive term, constant with respect to ${ \aa}$, may be ignored
since our concern is with entropy differences.
Note that ${ h_{\phi}=0}$ and that the notation ${ h_{\aa}}$ presumes
invariance to any permutation of the subscripts, justified because the
underlying distribution is invariant.
\par
In reporting numerical values of the entropy, or more usually differences
in entropy,  ${ h}$ is scaled to millibits  by multiplying by the factor
${ 2^{10}/\log(2)}$. 
The upper limit for the mutual information against independence for
two binary variables with equi-probable margins is ${ 1024}$mbits,
attained when the variables always take the same value.
For two Gaussian variables with correlation ${ 0.5}$ the measure is
${ 212.5}$mbits, but there is no upper limit.  
\par 
For disjoint sets ${ \aa,\bb,\cc\in\calP}$ the conditional mutual
information is
{\begin{eqnarray}\label{eq:cmih}
\I_{ \aa\bb|\cc}=-h_{\aa\cup\bb\cup\cc}+h_{\aa\cup\cc}+h_{\bb\cup\cc}-h_{\cc}.
\end{eqnarray}}\blank\newline
It is useful to retain both the ${ \inf}$ and ${ \I}$ notations for this measure.
The marginal information is ${ \I_{ ij}}$ where the conditioning set is empty.
We require the well known lemma that
{\begin{eqnarray}\label{eq:cmilemma}
\I_{ \aa\bb|\cc}=0 \iff X_{\aa}\indep{}X_{\bb}\mid{}X_{\cc}.
\end{eqnarray}}\blank\newline
The proof uses the non-negativity of the Kullback-Liebler divergence
between the joint distribution and the distribution factorised
according to the independence statement.
\par
\subsection{\sffamily Entropy function expansion }
The entropy function ${ h}$ is defined on the power set of
the nodes, ${ \{h_{\aa};\aa\in\calP\}}$.
The forward differences ${ \{\delta_{ \aa};\aa\in\calP\}}$ of the entropy
are defined by the additivity relations
{\begin{eqnarray}\label{eq:fwddiff}
h_{\aa} &=& \sum_{\bb\subseteq{}\aa} \delta_{ \bb} \jfor{} \aa\in\calP.
\end{eqnarray}}\blank\newline
Solving by Mobius inversion,  \cite{rota1964foundations}, gives
{\begin{eqnarray}\label{eq:inversion}
\delta_{ \aa} &=& \sum_{\bb\subseteq{}\aa} 
(-1)^{|\aa|-|\bb|}h_{\bb}\jfor{} \aa\in\calP.
\end{eqnarray}}\blank\newline
\begin{thm}\label{thm:symm}{\em\sf Symmetry of the forward differences.}
The forward differences are symmetric, that is, any ${ \delta_{ \aa}}$ is
invariant to any permutation of the indices within the set ${ \aa\in\calP}$.
\end{thm}
A  detailed proof is given in  the Appendix.
\par
We need the following lemma in a later section.
\begin{lmm}\label{lemm:addit}{\em\sf Additivity.}
When the entropy  is additive, so that 
${ h_{a\cup{}b}=h_{a}+h_{b}}$ 
for all 
${ a\subseteq\aa}$, ${ b\subseteq\bb}$, 
non-empty ${ a}$ and ${ b}$, 
and disjoint ${ \aa,\bb\subseteq\pp}$,
then ${ \delta_{ a\cup{}b}=0}$.
The converse also holds.
\end{lmm} 
The proof, given in the Appendix, essentially invokes
a triangular elimination scheme.
\par
\subsection{\sffamily Conditional mutual information  and forward differences}
From (\ref{eq:cmih}) the conditional mutual information of two random
variables ${ X_i,X_j}$ given a subset of others, ${ X_{\aa}}$, is
{\begin{eqnarray}\label{eq:cmi2}
\I_{ ij|\aa} &=& -h_{\aa{}ij}+h_{\aa{}i}+h_{\aa{}j}-h_{\aa},
\end{eqnarray}}\blank\newline
where ${ \aa{}ij}$ is shorthand for ${ \aa\cup\{i,j\}}$.
The right hand side is the elementary imset representation,
for pairwise conditional independence, \cite{studeny2005probabilistic}.
It is the scalar product of the entropy function with the imset
(integer valued multi-set) ${ (\dots,-1,1,1,-1\dots)}$ of length ${ 2^p}$
that has zeros in the appropriate places.
It is elementary because it represents a single conditional
independence statement.
\begin{thm}\label{thm:cmi}
{\em\sf Conditional mutual information and forward differences.}
The conditional mutual information can be expressed in terms of the  forward
differences, ${ \{\delta\}}$, of the entropy function by
{\begin{eqnarray}\label{eq:cmidelta}
 \I_{ ij|\aa} &=& - \sum_{ij\subseteq{}\bb\subseteq{}\aa{}ij}\delta_{ \bb} 
\jfor{}i,j\in\pp,\aa\in\calP.
\end{eqnarray}}\blank\newline
\end{thm}
\jpar{Remarks:}
The subset ${ \{i,j\}}$ occurs in every term on the right of (\ref{eq:cmidelta}).
The first term is the marginal mutual information ${ \I_{ ij}}$.
Each ${ \delta}$ term on the right is invariant to
permutation of its suffices.
If the conditioning set ${ \aa}$ is of moderate size then there are only
a moderate number of terms in the summation.
\begin{coroll}\label{coroll:3rdorder}{\em\sf Third order  forward differences.}\label{sect:coroll}
When ${ \aa=\{k\}}$ consists of a single element
{\begin{eqnarray}
    \delta_{ ijk}  &=& \I_{ ij}-\I_{ ij|k} ,\jand{}\label{eq:delta2}\\
 &=& 
h_{ijk}- h_{ij}-h_{ik}-h_{jk}
 +h_{i}+h_{j}+h_{k}-h_{\phi}.\label{eq:deltah}
\end{eqnarray}}\blank\newline
\end{coroll}
This follows because setting ${ \aa=\phi}$ in (\ref{eq:cmidelta}) gives
${ \I_{ ij}=-\delta_{ ij}}$, and ${ \aa=\{k\}}$ gives
${ \I_{ ij|k}=-(\delta_{ ij}+\delta_{ ijk})}$.
Subtraction gives (\ref{eq:delta2}).
The second statement is just the inversion formula
(\ref{eq:inversion}) for ${ \delta_{ ijk}}$.
\par
This corollary locates  the identity introduced at (\ref{eq:delta})
within a wider framework.
The key property is the difference ${ \delta}$ is symmetric in
permutation of suffices ${ i,j,k}$, as in (\ref{eq:deltah}) while
intuitively the right hand side of (\ref{eq:delta2}) is not.
\par
\subsection{\sffamily Forward differences of the conditional entropy}
The conditional entropy function
${ \{h_{\aa|\bb};\aa\in\calP(\pp\exc\bb)\}}$ is defined on the restricted
power set that excludes ${ \bb}$ where
${ h_{\aa|\bb}=-\E\log\f_{ \aa|\bb}(X_{\aa}|X_{\bb})}$.
The corresponding conditional forward differences are 
defined by (\ref{eq:fwddiff}) and (\ref{eq:inversion})
giving ${ \{\delta_{ \aa|\bb};\aa\in\calP(\pp\exc\bb)\}}$.
The set notation in ${ \delta_{ \aa|\bb}}$ makes evident the symmetry of 
the differences.
\begin{thm}\label{thm:fwdident}{\em\sf A recursion 
for conditional forward differences.}
For ${ k\in\pp}$,
${ \bb\subseteq{}\pp\exc{}k}$ and ${ \aa\in\calP(\pp\exc{}(\bb{}\cup{}k))}$
the conditional forward differences satisfy
{\begin{eqnarray}\label{eq:recursion}
 \delta_{ \aa{}|\bb{}k} &=&  \delta_{ \aa{}k|\bb{}}+\delta_{ \aa|\bb{}{}}.
\end{eqnarray}}\blank\newline
\end{thm}
\jpar{Remarks:}
When ${ \bb}$ is empty, the identity (\ref{eq:recursion}) shows that the
higher order forward difference is the difference between a
conditional and a marginal forward difference:
{\begin{eqnarray*}
\delta_{ \aa{}k}=\delta_{ \aa{}|k}-\delta_{ \aa{}}.
\end{eqnarray*}}\blank
The size of the higher order term ${ \delta_{ \aa{}k}}$ is useful in
assessing how much ${ \delta_{ \aa{}}}$ might change by conditioning on a
further variable.
This is invariant to permutation of the set ${ \aa{}k\equiv\aa\cup{}\{k\}}$.
To illustrate with ${ |A|=3}$,  
${ \delta_{ 1234} = \delta_{ 123|4}-\delta_{ 123} = \delta_{ 124|3}-\delta_{ 124}}$
and so on.
\par
The identity (\ref{eq:recursion})  generalises
to express a conditional forward
difference  as sums of conditional forward differences
conditioning on a lower order:
{\begin{eqnarray*} \delta_{ \aa{}|\bb\cup\cc} &=&  \sum_{\dd\subseteq\cc}\delta_{ \aa{}\cup\dd|\bb}.
\end{eqnarray*}}\blank
\par
\begin{thm}\label{thm:sepfwd}{\em\sf Separation and the forward difference.}
Whenever ${ \cc}$ separates ${ \aa}$ and ${ \bb}$ in the conditional independence graph
${ \delta_{ \aa\cup\bb|\cc}=0}$.
\end{thm}
\jpar{Remarks:} 
The value of this result is that it allows easy
interpretations of marginal forward differences in examples.
There is a converse to this theorem if the condition on the 
conditional forward differences is strengthened.
\par
\subsection{\sffamily Non-collapsibility of mutual information}
Collapsibility is important in statistical inference because 
it elucidates which properties of a joint distribution can be 
inferred from a margin.
Simpson's paradox \cite{simpson1951interpretation} refers to a
violation of collapsibility; other references are
\cite{bishop1975discrete},
\cite{whittemore1978collapsibility}, 
\cite{whittaker1990graphical}, \cite{greenland1999confounding}, among
others.
\par
Consider three variables with  ${ I_{ ik|j}=0}$ and corresponding  
independence  graph
\par\begin{center}
{\small\begin{picture}(100,20)(0,-10)
 \put(0,0){{\circle{20}}\makebox(0,0){${ i}$}}
\put(50,0){{\circle{20}}\makebox(0,0){${ j}$}}
\put(100,0){{\circle{20}}\makebox(0,0){${ k}$}}
 \put(  10,0 ) {\line(1,0){30}}
 \put(  60,0 ) {\line(1,0){30}}
\end{picture}}
\end{center}
with one missing edge.
The strength of the relationship between ${ X_{i}}$ and ${ X_{j}}$ is
measured in two dimensions by ${ I_{ ij}}$ and in three dimensions by
${ I_{ ij|k}}$.
If ${ I}$ were collapsible then ${ I_{ ij}-I_{ ij|k}=0}$.
But this difference is ${ -\delta_{ ij}+\delta_{ ij|k}=\delta_{ ijk}}$
by (\ref{eq:delta2}), the 3rd-order difference.
By symmetry ${ \delta_{ ijk}}$ is also equal to ${ \delta_{ ik|j}-\delta_{ ik}}$
and so ${ \delta_{ ijk}=0}$ together with ${ \delta_{ ik|j}=0}$
 would imply ${ \delta_{ ik}=0}$;
which is false in general.
The premiss that the measures are equal is untenable.
\jpar{Remarks:} 
Large values of ${ \delta_{ ijk}}$ indicate that conditioning on ${ X_{k}}$
modifies the strength of the relationship between ${ X_{i}}$ and ${ X_{j}}$;
even though it is a symmetric measure this does not imply that that
subgraph be complete.
\par
More generally requiring the collapsibility of ${ \delta_{ \aa}}$ in the
space ${ \aa\cup\bb}$ requires ${ \delta_{ \aa|\bb}=\delta_{ \aa}}$; by
Theorem \ref{thm:sepfwd} this is equivalent to
${ X_{\aa}\indep{}X_{\bb}}$.
\par
\section{\sffamily\large Synergy, suppression and immorality}
\par
\subsection{\sffamily Synergy}
The information against the independence of two variables is
synonymous with the information explained in one variable by
predicting from the other.
\begin{thm}\label{thm:synergy}{\em\sf Explained information.}
The explained information in one variable expressed in terms of the
marginal mutual information of others is
{\begin{eqnarray}\label{eq:general}
\inf(X_{k}\indep{}X_{\aa})  &=&
\sum_{i\in\aa}\inf(X_{k}\indep{}X_{i}) 
- 
\sum_{\bb\subseteq\aa,\mid{}\bb|>1}\delta_{ \bb{}k},
\end{eqnarray}}\blank\newline
where the last summation is over subsets ${ \bb}$ with at least ${ 2}$ elements.
In particular
{\begin{eqnarray}\label{eq:synergy}
\inf(X_{k}\indep{}(X_{i},X_{j}))  &=&
\inf(X_{k}\indep{}X_{i}) + \inf(X_{k}\indep{}X_{j}) - \delta_{ ijk}.
\end{eqnarray}}\blank\newline
\end{thm}
\par
The proof is included in the Appendix.
When ${ \delta_{ ijk}<0}$ the triple ${ \{i,j,k\}}$ is called a
{\em~synergy}, as the total information explained exceeds the sum of
the marginal informations taken alone.
It is appropriate to label the triple a synergy, rather than the
variable ${ k}$, since (\ref{eq:synergy}) is invariant to permutation of
the indices.
\par
\begin{coroll}\label{coroll:partial}{\em\sf Partially explained information.}
The explained information in one variable expressed in terms of the
marginal mutual informations  of variables in ${ \aa}$ adjusted for
variables in ${ \bb}$ is
{\begin{eqnarray}\label{eq:general}
\inf(X_{k}\indep{}X_{\aa}\mid{}X_{\bb})  &=&
\sum_{i\in\aa}\inf(X_{k}\indep{}X_{i}\mid{}X_{\bb}) 
- 
\sum_{\cc\subseteq\aa,|\cc|>1}\delta_{ \cc{}k|\bb}.
\end{eqnarray}}\blank\newline
When there are just two variables in ${ \aa}$
{\begin{eqnarray}\label{eq:general}
\inf(X_{k}\indep{}(X_{i},X_{j})\mid{}X_{\bb})  &=&
\inf(X_{k}\indep{}X_{i}\mid{}X_{\bb}) + \inf(X_{k}\indep{}X_{j}\mid{}X_{\bb}) 
- \delta_{ ijk|\bb},
\end{eqnarray}}\blank\newline
the sum of the parts adjusted by the conditional 3rd-order difference.
\end{coroll}
The proof follows the previous argument  and is straightforward.
When ${ \delta_{ ijk|\bb}<0}$ the triple ${ \{i,j,k\}}$ is also called a
synergy though a conditional or partial synergy is more specific.
\par
\subsection{\sffamily Suppression}
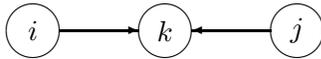
\begin{figure}[!hbt]
\begin{center}
{\small\begin{picture}(100,20)(0,-10)
\put(0,0){{\circle{20}}\makebox(0,0){${ i}$}}
\put(50,0){{\circle{20}}\makebox(0,0){${ k}$}}
\put(100,0){{\circle{20}}\makebox(0,0){${ j}$}}
 \put(  10,0 ) {\vector(1,0){30}}
 \put(  90,0 ) {\vector(-1,0){30}}
\end{picture}}
\end{center}
\caption{Suppression and immorality; in a supressor regression context
${ j}$ is the dependent variable, ${ k}$ the explanatory
variable, and ${ i}$ the supressor variable.\label{fig:immor}}
\end{figure}

The term suppressor variable is used in regression
applications where there is an contextual asymmetry between the
dependent and the explanatory variable,
see the introductory section.
The suppressor variable describes a third variable which is
uncorrelated (or nearly so) with the dependent variable, but adds to
the predictive power of the regression on both; this is technically
described by ${ \delta_{ ijk}<0}$ from (\ref{eq:synergy}) together with
${ I_{ ij}=0}$.
The corresponding  Bayes network  is displayed in Figure \ref{fig:immor}.
\par
The diagram makes clear that suppression is symmetric in the sense
that the variables ${ i}$ and ${ j}$ are interchangeable.
Elaboration of this condition in terms of correlations is 
the content of Theorem  \ref{sect:3dcor}.1
in the Applications Section.
\jpar{Remarks:} Expressing the criterion in the more general framework
of information theory, extends the idea of suppression in linear
regression to variables measured on other scales with well defined
information measures, including categorical data.
Examples are  given in the next section. 
\par{}
Expressing suppression in the general terms of information, 
clarifies the issues when more than two explanatory
variables are involved.
Recognition of a synergy in a particular context could just reduce to
calculating the conditional 3rd-order difference ${ \delta_{ ijk|\bb}}$.
Screening for synergies or partial synergies involves repeated
calculations, there are many triples as ways of choosing the two
explanatory variables from the candidate set.
\par 
An alternative direction is to develop (\ref{eq:general}).  
For instance with three explanatory variables and variable ${ k=1}$
dependent the synergy criterion becomes
{\begin{eqnarray*}
\delta_{ 123}+\delta_{ 124}+\delta_{ 134}+\delta_{ 1234}<0.
\end{eqnarray*}}\blank
If, as well, only ${ \delta_{ 1234}<0}$ suppression is truly a function of
the three explanatory variables taken as a whole.
\par
\subsection{\sffamily Immorality}
The concept of an unshielded collider, \cite{pearl1988probabilistic},
is key for understanding d-separation in Bayes networks
on acyclic directed graphs.
\cite{lauritzen1988local} refer to the same concept as an immorality.
Bayes networks with different directions may be probabilistically
equivalent and an important result in this area,
\cite{frydenberg1990chain,verma1990equivalence}, is that the
equivalence class is characterized by the skeleton of the graph and
its unshielded colliders.
An unshielded collider is displayed in Figure \ref{fig:immor} for
three variables, where the absence of an arrow joining ${ i}$ and ${ j}$
indicates ${ X_{i}\indep{}X_{j}}$.
Consequently ${ I_{ ij}=0}$, but ${ I_{ ij\mid{}k}>0}$ so that ${ \delta_{ ijk}<0}$ by
(\ref{eq:delta2}), the same condition as for suppression.
\par
In a Bayes network with additional antecedent variables the
condition for an unshielded collider requires that ${ k}$ of Figure
\ref{fig:immor} is not in the separation set ${ \aa}$ for which
${ I_{ ij|\aa}=0}$.
This translates to 
\begin{thm}\label{thm:collider}{\em\sf  Characterization 
of an unshielded collider.}
In a Bayes network the condition
{\begin{eqnarray}\label{eq:immor}
I_{ ij|\aa}=0\jand\delta_{ ijk|\aa}\leq{}0
\end{eqnarray}}\blank\newline
where ${ \delta_{ ijk|\aa}}$ is the 3rd-order conditional forward difference
is necessary and sufficient for an unshielded collider at ${ k}$. 
\end{thm}\par
The proof is just a rephrasing of the definition of unshielded collider.
\jpar{Remarks:} This result gives an interpretation for negative conditional
3rd-order forward differences, and suggests a method of identifying
immoralities in a Bayes network.
\par
\subsection{\sffamily Systematic computation of low order forward differences  }
The forward differences of the entropy offer low dimensional summaries of
the data. 
Consider which differences to compute.
For a small number of variables, all are possible, but for moderate
and large numbers this is a formidable task.
Furthermore the differences required may be different for the analysis
of regression and suppression, for contingency table analysis and
collapsibility, and for direction determination in Bayes networks.
\par
In the analysis of a candidate graph second order differences are
routinely computed as marginal informations for all pairs.
Third order differences complement the information from the nested
pairs and may flag suppression and immorality, which are interesting
because they are infrequent.
We propose these are computed for a subset of the triples of a given
graph.
Fourth order difference show changes in third order differences and
hence may flag conditional synergies; 
we suggest these are only computed in relation to specific triples of
interest.
\par
{\em\sf Requirements:}
The subset of triples are required to cover the graph without unnecessary
computing.
In particular previously computed differences should not be recomputed
nor should redundant ones that have an priori zero value with respect
to the graph.
Because higher order conditional mutual informations are additive in
lower order differences, see (\ref{eq:cmidelta}), it is desirable to require
{\em\sf nested} subsets, so that for example if a fourth order difference
is computed its corresponding lower order differences are available.
\par{\em\sf Redundancy:}
For a given conditional independence graph certain forward differences
are either identically zero, or reduce to linear combinations of lower
order differences.
For instance if ${ i}$ and ${ j}$ belong to separate connected components of
the graph  ${ \delta_{ \aa}=0}$ whenever ${ \aa}$ includes both ${ i}$ and
${ j}$.
Consequently when a putative graph describing the dependencies in the
data is given, not all forward differences are interesting.
Note that ${ \delta_{ ijk}=0}$ whenever the subgraph of the triple is not connected.
The proof is straightforward: the subgraph is not connected when
${ \inf(X_{k}\indep{}(X_{i},X_{j}))=0}$.
Consequently both ${ \I_{ ik}=0}$ and ${ \I_{ jk}=0}$ and so
${ \delta_{ ijk}=0}$ by (\ref{eq:synergy}).
Only connected triples have interesting forward differences.
Restricting attention to subsets that have complete subgraphs with
respect to a given graph satisfies the nesting criterion, 
but would disallow  computation of a third order difference
on the chain in Figure \ref{fig:immor}.
\par{\em\sf Node clusters}:
We suggest that a subset of nodes in which one node has an edge
to every other node is a configuration for which it is appropriate
to compute forward differences of an  order up to the subset size.
Node clusters of this form have an approximate nesting structure: all
but one subsets of the cluster are clusters themselves, and if
any one edge is dropped from the cluster, it leaves a cluster.
The complete subgraph on any number of nodes is a node cluster where
any one of the vertices may take the role of the cluster node.
Certain configurations are eliminated, for instance, a chain or a
chordless cycle on four variables.
A subset of size ${ 3}$ forms a node cluster if one node is adjacent to both
others.
\par 
An algorithm based on this concept is included in the Supplementary
Material. 
\par{\em\sf Collider colouring of the synergies}:
Synergies are infrequent and so of interest.
They are a property of a triple and so more difficult to portray than
a node.
However the node opposite the weakest edge may be singled out as a
collider, generalising the term used in Bayes networks, 
\cite{pearl1988probabilistic}.
When the weakest edge has zero mutual information, so that its nodes
are marginally independent, then the resulting configuration is an
unshielded collider (immorality) and the two notions coincide.
\par 
The collider may be indicated by a colour (red, say) and the other two
nodes yellow, 
Colouring the two edges adjacent to the collider indicate the other
elements of the triple.
Additional rules are needed for overlapping synergies; for instance,
any node tagged as a collider is overprinted as a collider.
This can  lose some detail of the synergy in the graph.
\par
\section{\sffamily\large Applications of forward  differences }  
We give some low dimensional examples of forward differences of the
entropy, both theoretical and empirical, for categorical and
continuous data.
In three dimensions the third order differences quantify the
difference in mutual information between two variables with and
without conditioning on a third.
We compute and display these differences from some known standard
models numerically and, where possible,  give an analytic condition
for a synergy.
A difference is measured in millibits, the same units that measure entropy.
For continuous data, we elaborate the conditions for suppression for 
 a theoretical variance matrix with a known graph structure, and give  
some simple examples.
For categorical data we illustrate synergy with examples of 
binary data in  three dimensions, and relate these to the issue of 
collapsibility.
We elucidate examples of  four dimensions continuous models that are
interesting in the context of Bayes networks.
\par
Higher dimensional examples discuss 3rd-order forward differences and
synergies using the skeleton of the Bayes network, known or postulated to
have generated the data.
Firstly from an artificial tree averaging process, which establishes
why the skeleton rather than the moral graph is the right graph to
determine which differences need to be computed.
Secondly the real-life example of wine quality data is analysed and
the synergies suggest that a chain graph model might represent the
structure of the variables well.
\bNotBmka 
The analysis of the  carcass data leads to similar conclusions, 
but is included because it is easily accessible through R.
\eNotBmka 
\par
\subsection{\sffamily Three dimensional correlations}\label{sect:3dcor}
The lower off-diagonal elements of the correlation matrix ${ \Sigma}$
are ${ \rho_{12},\rho_{13},\rho_{23}}$, constrained by requiring ${ \Sigma}$
to be positive definite.
The Gaussian entropy function is given in the preliminary remarks
to Section \ref{sect:fwddiff}.
The power set has ${ 2^3}$ elements, the entropy of the singleton sets
are standardised to zero and all others are negative.
The 2nd-order forward differences are negatives of the marginal mutual
informations, so that the information against the independence of
${ X_{1}}$ and ${ X_{2}}$ is ${ \delta_{ 12}=-I_{ 12}=\log(1-\rho_{12}^2)/2}$.
\begin{thm}\label{thm:gauss}{\em\sf Synergy with three Gaussian variables.}
The 3rd-order forward difference is
{\begin{eqnarray}\label{eq:d123gauss1}
\delta_{ 123}&=&\half\log
\frac{1-\rho_{12}^2-\rho_{13}^2-\rho_{23}^2+2\rho_{12}\rho_{13}\rho_{23}}
     {(1-\rho_{12}^2)(1-\rho_{13}^2)(1-\rho_{23}^2)},\\
&=&\half\log\frac{1-\rho_{12|3}^2}{1-\rho_{12}^2},\label{eq:d123gauss2}
\end{eqnarray}}\blank\newline
and the condition for a synergy, ${ \delta_{ 123}<0}$, is that one
marginal correlation coefficient is smaller than its corresponding
partial in absolute value, for instance ${ |\rho_{12}|<|\rho_{12|3}|}$.
\end{thm}
The proof is in the Appendix.
\begin{coroll}{\em\sf Synergy and negative correlation.}
A synergy occurs whenever exactly one  marginal correlation is
negative.
\end{coroll}
\jpar{Remarks:}
There are only two  cases of correlation matrix to consider: one
where all coefficients are positive and the other where exactly one
is negative. 
The corollary deals with the second.
It follows because if one correlation is negative then the
corresponding partial, say
${ \rho_{12|3}=(\rho_{12}-\rho_{13}\rho_{23})
\{(1-\rho_{13}^2)(1-\rho_{23}^2)\}^{-\half}}$, exceeds the marginal in
terms of absolute value since the numerator is inflated and the
denominator deflated.
\par
In regression scenarios the condition that one marginal correlation is
negative may be subdivided by whether the correlation is between a
response and an explanatory variable, or between two explanatory
variables.
This corresponds to the classification of suppression into type:
negative or reciprocal, occuring in the literature on suppression
and briefly reviewed in the Introduction.
The special case ${ \rho_{12}=0}$ corresponds to classical suppression.
\par
Of interest to us is that a synergy does not occur when
${ \rho_{12|3}=0}$, and the inequality condition
${ |\rho_{12}|<|\rho_{12|3}|}$ is invariant to permuting indices.
\par
\begin{example} (Numerical):
For a numerical illustration the
 forward differences are displayed using ${ \Sigma}$ specified by its
lower triangle ${  1.0, 0.2, 1.0, 0.7, 0.5, 1.0}$, and for comparison, of the same
${ \Sigma}$ with ${ 0.2}$ replaced by ${ -0.2}$.
The  forward differences are, respectively,
\begin{center}{\small\begin{tabular}{r | rrrr rrrr}
subset&${ \phi}$&1&2&3&12&13&23&123\\ \hline
fwd.diff(${ \rho_{12}=0.2}$) &0&0&0&0&-30.15&-497.4&-212.5&-14.63\\ 
fwd.diff(${ \rho_{12}=-0.2}$)&0&0&0&0&-30.15&-497.4&-212.5&-1126.0
\end{tabular}}\end{center}
\par
The values are reported in millibits, see the preliminaries to 
Section \ref{sect:fwddiff}.
In both the information against ${ X_{1}\indep{}X_{2}}$ is ${ 30.15}$mbits.
In the first instance the 3rd-order difference ${ \delta_{ 123}}$ is
${ -14.63}$mbits so that the information against
${ X_{1}\indep{}X_{2}\mid{}X_{3}}$ is ${ 30.15+14.63=44.78}$mbits.
In the second instance ${ \delta_{ 123}=-1126}$mbits indicating a much more
substantial synergy.
\end{example}
\par
The result (\ref{eq:d123gauss2}) generalises easily to give a condition
for partial synergy.
\begin{coroll}{\em\sf Partial synergy with three Gaussian variables.}
The 3rd-order conditional forward difference for three
Gaussian variables given a set ${ \aa}$ of other such variables is
{\begin{eqnarray}
\delta_{ 123|\aa}
&=&\half\log\frac{1-\rho_{12|\aa3}^2}{1-\rho_{12|\aa}^2}\label{eq:d123gauss3}
\end{eqnarray}}\blank\newline
and the condition for a partial synergy, ${ \delta_{ 123|\aa}<0}$, is that one
marginal correlation coefficient is smaller than its corresponding
partial in absolute value, that is ${ |\rho_{12|\aa}|<|\rho_{12|\aa3}|}$.
\end{coroll}
\par
\subsection{\sffamily Three dimensional contingency tables}\label{sect:3dtab}
While the value of the 3rd-order difference clearly flags the
phenomenon of suppression in regression it does not give a definitive
answer to non-collapsibility in three way tables.
We consider three examples related to Simpsons paradox: the first is
an archetypal loglinear model, the second is numerical and the third
is real-life.
\par
\begin{example} (Analytic):
The first example of a ${ 2^3}$-table is analytic where each margin shows
independence but the three variables are dependent.
A priori the 3rd-order forward difference must be negative.
\par
The log-linear is expansion of ${ p_{ 123}}$ on ${ \{0,1\}^3}$ is
{\begin{eqnarray}
\log(\alpha)+
    (x_1+x_2+x_3)\log(\beta/\alpha)-
    2(x_1x_2+x_1x_2+x_2x_3)\log(\beta/\alpha)+4x_1x_2x_3\log(\beta/\alpha),
\end{eqnarray}}\blank\newline
where ${ x_1,x_2,x_3}$  take values ${ 0,1}$; and parameterised by 
${ \alpha\in(0,1/4)}$ with  ${ \beta=1/4-\alpha}$.
In standard order the joint probabilities are
${ (\alpha,\beta,\beta,\alpha,\beta,\alpha,\alpha,\beta)}$.
It illustrates non-collapsibility because every margin has
equi-probability entries so that ${ \inf(X_{i}\indep{}X_{j})=0}$,
while any two variables contribute positively to the prediction
of the third.
\par
 By direct evaluation,
{\begin{eqnarray*}
\delta_{ 123}=-4(\alpha\log(\alpha)+\beta\log(\beta))-3\log(2).
\end{eqnarray*}}\blank
This is zero when ${ \beta=\alpha}$, but otherwise negative.
\end{example}
\par
\par
\begin{example} (Kidney stones):
This is taken from \cite{julious1994confounding} has previously
been used as a real-life instance of Simpson's paradox.
There are two factors (Treatment, Size), each with two levels (A/B,
small/large stones respectively).
Outcomes (81/87, 234/270, 192/263, 55/80) are recorded as the
success/total count in the four groups, in Treatment within Size
order.
\par
The entropy function and its forward differences are displayed here
\begin{center}
{\small\begin{tabular}{r | rrrr rrrr}
subset&${ \phi}$&O&T&S&OT&OS&TS&OTS\\ \hline
entropy &0.0&733.4&1024.0&1023.7&1754.9&1725.8&1835.3&2533.7\\ 
fwd.diff.~${ \delta}$ &0&733.4&1024.0&1023.7&-2.443&-31.31&-212.4&-1.198\\
\end{tabular}}
\end{center}
with values in millibits.
The ${ T}$ margin is exactly balanced (1024mbits is the maximum),
and the ${ S}$ margin almost so, but the ${ T\times{}S}$ table is not
(the mutual information is ${ 212.4}$mbits and far from zero).
The value of ${ \delta_{ OTS}=-1.198}$mbits is negative; it is also
negligible so that marginal and conditional independence measures are
approximately the same.
The independence graph approximating these data is 
\begin{center}
{\small\begin{picture}(100,20)(0,-10)
\put(0,0){{\circle{20}}\makebox(0,0){${ O}$}}
\put(50,0){{\circle{20}}\makebox(0,0){${ S}$}}
\put(100,0){{\circle{20}}\makebox(0,0){${ T}$}}
 \put(  10,0 ) {\line(1,0){30}}
 \put(  90,0 ) {\line(-1,0){30}}
\end{picture}}
\end{center}
\par
Here Simpson's paradox occurs when comparing the ${ OT}$ interaction conditionally on ${ S}$, with its value marginalised over ${ S}$, and arises
because of the large imbalance in the ${ T}$x${ S}$ table.
The value of ${ \delta_{ OTS}}$ does not signal the paradox.
\end{example}
\par
It is easy to construct examples where the paradox (log odds ratio in
the marginal and in the conditional tables are of opposite sign) goes
with a negative and examples with a positive third order difference.
\par
\subsection{\sffamily Four dimensional correlation matrices}\label{sect:4dcor}
\begin{example} (Analytic):
The forward differences of the entropy are calculated from the 
theoretical correlation matrix of various  four dimensional graphical models.
We compute forward differences all orders, though report only the most
salient features to illustrate what may be expected if data is
generated from such models.  
The graphical models, characterized by the graphs in
Figure~\ref{fig:4dcor}, include the so-called cluster model, a chain,
a decomposable model, the 4-cycle, Bayes networks with one, two and
three unshielded colliders.  
\par 
The interpretation of their differences derives from the separation
properties of the graph translating to a statement of the form
${ \delta_{ \aa\cup\bb|\cc}=0}$ in Theorem \ref{thm:sepfwd}; this in turn
leads to one or more linear relationships using Theorem \ref{thm:fwdident}.
These results are summarised in Table \ref{tab:4dsummary}. 
\begin{figure}[!p]
\begin{center}{\tiny
\begin{picture}(400,180)(-20,-20)
 \unitlength .8pt             
 \put(0,80){\begin{picture}(100,50)(0,-10)             \put(0,40){\makebox(0,0){(a) cluster}}
           \put(0,0){{\circle{20}}\makebox(0,0){${ 1}$}}
           \put(50,0){{\circle{20}}\makebox(0,0){${ 2}$}}
           \put(100,0){{\circle{20}}\makebox(0,0){${ 3}$}}
           \put(50,50){{\circle{20}}\makebox(0,0){${ 4}$}}
           \put(  10,0 ) {\line(1,0){30}}
           \put(  50,10 ) {\line(0,1){30}}
          \put(  90,0 ) {\line(-1,0){30}}
 \end{picture}}
 \put(150,80){\begin{picture}(100,50)(0,-10)
           \put(0,40){\makebox(0,0){(b) chain}}
           \put(0,0){{\circle{20}}\makebox(0,0){${ 1}$}}
           \put(50,0){{\circle{20}}\makebox(0,0){${ 2}$}}
           \put(100,0){{\circle{20}}\makebox(0,0){${ 3}$}}
           \put(100,50){{\circle{20}}\makebox(0,0){${ 4}$}}
           \put(  10,0 ) {\line(1,0){30}}
           \put(  100,10 ) {\line(0,1){30}}
           \put(  90,0 ) {\line(-1,0){30}}
 \end{picture}}
 \put(300,80){\begin{picture}(100,50)(0,-10)
           \put(0,40){\makebox(0,0){(c) decomp}}
           \put(0,0){{\circle{20}}\makebox(0,0){${ 1}$}}
           \put(50,0){{\circle{20}}\makebox(0,0){${ 2}$}}
           \put(100,0){{\circle{20}}\makebox(0,0){${ 3}$}}
           \put(50,50){{\circle{20}}\makebox(0,0){${ 4}$}}
           \put(  10,0 ) {\line(1,0){30}}
           \put(  50,10 ) {\line(0,1){30}}
           \put(  90,0 ) {\line(-1,0){30}}
           \put(  57,43 ) {\line(1,-1){35}}
 \end{picture}}
 \put(470,80){\begin{picture}(100,50)(0,-10)
           \put(-30,40){\makebox(0,0){(d) 4-cycle}}
           \put(0,0){{\circle{20}}\makebox(0,0){${ 1}$}}
           \put(50,0){{\circle{20}}\makebox(0,0){${ 2}$}}
           \put(50,50){{\circle{20}}\makebox(0,0){${ 3}$}}
           \put(0,50){{\circle{20}}\makebox(0,0){${ 4}$}}
           \put(  10,0 ) {\line(1,0){30}}
           \put(  10,50 ) {\line(1,0){30}}
           \put(  0,10 ) {\line(0,1){30}}
           \put(  50,10 ) {\line(0,1){30}}
 \end{picture}}
                \put(40,-20){\begin{picture}(100,50)(0,-10)
           \put(-40,40){\makebox(0,0){(e) bayesNetA}}
           \put(0,0){{\circle{20}}\makebox(0,0){${ 1}$}}
           \put(50,0){{\circle{20}}\makebox(0,0){${ 2}$}}
           \put(50,50){{\circle{20}}\makebox(0,0){${ 3}$}}
           \put(0,50){{\circle{20}}\makebox(0,0){${ 4}$}}
           \put(  10,0 ) {\vector(1,0){30}}
           \put(  10,50 ) {\vector(1,0){30}}
           \put(  0,10 ) {\vector(0,1){30}}
           \put(  50,10 ) {\vector(0,1){30}}
 \end{picture}}
 \put(200,-20){\begin{picture}(100,50)(0,-10)
           \put(-40,40){\makebox(0,0){(f) bayesNetB}}
           \put(0,0){{\circle{20}}\makebox(0,0){${ 1}$}}
           \put(50,0){{\circle{20}}\makebox(0,0){${ 2}$}}
           \put(50,50){{\circle{20}}\makebox(0,0){${ 3}$}}
           \put(0,50){{\circle{20}}\makebox(0,0){${ 4}$}}
           \put(  40,0 ) {\vector(-1,0){30}}
           \put(  10,50 ) {\vector(1,0){30}}
           \put(  0,40 ) {\vector(0,-1){30}}
           \put(  50,10 ) {\vector(0,1){30}}
 \end{picture}}
  \put(360,-20){\begin{picture}(100,50)(0,-10)
           \put(-40,40){\makebox(0,0){(g) bayesNetC}}
           \put(0,0){{\circle{20}}\makebox(0,0){${ 1}$}}
           \put(50,0){{\circle{20}}\makebox(0,0){${ 2}$}}
           \put(100,0){{\circle{20}}\makebox(0,0){${ 3}$}}
           \put(50,50){{\circle{20}}\makebox(0,0){${ 4}$}}
           \put(  10,0 ) {\vector(1,0){30}}
           \put(  50,40 ) {\vector(0,-1){30}}
           \put(  90,0 ) {\vector(-1,0){30}}
 \end{picture}}
\end{picture}
}\end{center}               
\caption{Four dimensional configurations of
         independence graphs (undirected and directed). 
    \label{fig:4dcor}}
\end{figure}
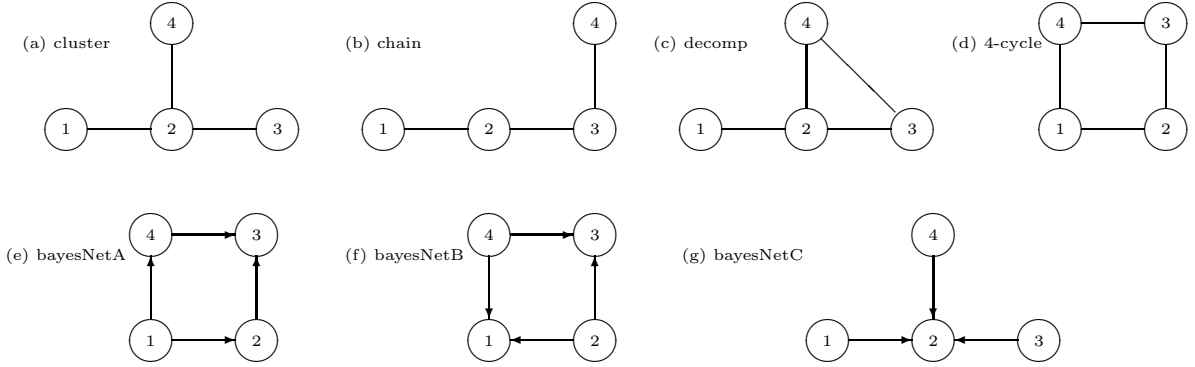
\begin{table}[!p]
\caption{Summary of four dimensional forward differences
         for examples  in Figure~\ref{fig:4dcor}.
     \label{tab:4dsummary}}
\begin{center}
${ {\begin{array}{ll}
 (a)\mbox{ cluster }
  &\delta_{ 13}+\delta_{ 123}=0,\quad \delta_{ 14}+\delta_{ 124}=0,
                   \quad  \delta_{ 34} + \delta_{ 234}=0,\\
  & \delta_{ 123}>0,\quad\delta_{ 124}>0,\quad\delta_{ 234}>0,\\
  & \delta_{ 134}>0,\quad\delta_{ 134} + \delta_{ 1234}=0, \quad\delta_{ 1234}<0.\\
 (b)\mbox{ chain }
  &
\delta_{ 13}+\delta_{ 123}=0,\quad 
\delta_{ 14}+\delta_{ 124}=0,\quad 
\delta_{ 14}+\delta_{ 134}=0,\quad 
\delta_{ 24}+\delta_{ 234}=0,\\
& \delta_{ 123}>0,\quad\delta_{ 124}>0,\quad\delta_{ 134}>0,\quad\delta_{ 234}>0,\\
& \delta_{ 14}+\delta_{ 124}+\delta_{ 134} +\delta_{ 1234}=0,\quad  \delta_{ 1234}<0.\\
(c)\mbox{ decomp }
&
\delta_{ 13}+\delta_{ 123}=0,\quad 
\delta_{ 14}+\delta_{ 124}=0,\quad\delta_{ 123}>0,\quad\delta_{ 124}>0,\\
& \delta_{ 134} +\delta_{ 1234}=0, \quad\delta_{ 1234}\mbox{ arbitrary. }\\
(d)\mbox{ 4-cycle }
&\delta_{ 13}+\delta_{ 123}+\delta_{ 134}+\delta_{ 1234}=0,
 \quad\delta_{ 24}+\delta_{ 123}+\delta_{ 124}+\delta_{ 1234}=0,\\
& \delta_{ 123|4}>0,\quad\delta_{ 124|3}>0,\quad\delta_{ 134|2}>0,
\quad\delta_{ 234|1}>0.\\
(e)\mbox{ bayesNetA }
&\delta_{ 24}+\delta_{ 124}=0, \quad
 \delta_{ 13}+\delta_{ 123}+\delta_{ 134}+\delta_{ 1234}=0,\\
  &  \delta_{ 124}>0, \quad\delta_{ 234|1}<0, 
\mbox{ other }\delta\mbox{s arbitrary.}\\
(f)\mbox{ bayesNetB }
&\delta_{ 24}=0, \quad
 \delta_{ 13}+\delta_{ 123}+\delta_{ 134}+\delta_{ 1234}=0,\\
  &  \delta_{ 124}<0, \quad\delta_{ 234}<0, 
\mbox{ other }\delta\mbox{s arbitrary.}\\
 (g)\mbox{ bayesNetC }
  & \delta_{ 13}=0,\quad\delta_{ 14}=0,\quad\delta_{ 34}=0,\quad\delta_{ 134}=0,\\
  & \delta_{ 123}<0,\quad\delta_{ 124}<0,\quad\delta_{ 234}<0.\\
\end{array}}}$\end{center}
\end{table}
\par
\par{(a) Cluster}:
This node cluster is a sparse configuration sufficiently complex 
that ${ \delta_{ 1234}}$ is not zero.
The  variables ${ X_{1},X_{3},X_{4}}$ are mutually independent given the
cluster node ${ X_{2}}$, consequently
three of the four 3rd-order differences involving the cluster node ${ X_{2}}$
are positive as the information conditioned on ${ X_{2}}$ is zero.
The term ${ \delta_{ 134}}$ is necessarily positive, for instance, because
${ X_{3}\indep{}X_{4}\mid{}X_{2}}$, and ${ X_{1}}$ is a predictor of ${ X_{2}}$,
so that ${ I_{ 34}>{}I_{ 34|1}>{}I_{ 34|2}=0}$ 
(or equivalently ${ \delta_{ 34}<{}\delta_{ 34|1}<{}\delta_{ 34|2}=0}$).
As ${ 0=\delta_{ 134|2}=\delta_{ 134}+\delta_{ 1234}}$, ${ \delta_{ 1234}}$ is
always negative.
\par{(b) Chain}:
${ X_{1}\indep{}X_{3}\mid{}X_{2}}$ implies ${ \delta_{ 123}>0}$,
similarly all other triples have a positive forward difference.
That ${ X_1\indep{}X_{34}\mid{}X_{2}}$ implies ${ \delta_{ 134|2}=0}$; this,
together with the identity ${ \delta_{ 134|2}=\delta_{ 134}+\delta_{ 1234}}$
involving the fourth order difference, implies ${ \delta_{ 1234}<0}$.
Th dependence structure of this graph is characterized by the values of 
${ \{\delta_{ 12},\delta_{ 23},\delta_{ 34},\delta_{ 123},\delta_{ 234}\}}$.
\par{(c) Decomposable}:
${ X_{1}\indep{}X_{3}\mid{}X_{2}}$ implies ${ \delta_{ 123}>0}$, similarly
${ \delta_{ 124}>0}$.
Because ${ \delta_{ 134|2}=0 = \delta_{ 134} + \delta_{ 1234}}$ 
they are of opposite sign, but otherwise arbitrary.
\par{(d) 4-cycle}:
There are two independences leading to 
two zero linear combinations:
${ X_{1}\indep{}X_{3}\mid{}X_{24}}$ translates to 
${ \delta_{ 13|24}=0}$ ${ =\delta_{ 13}+\delta_{ 123}+\delta_{ 134}+\delta_{ 1234}}$,
and
${ X_{2}\indep{}X_{4}\mid{}X_{13}}$ translates to 
${ \delta_{ 24|13}=}$ ${ 0}$ ${ =\delta_{ 24}+\delta_{ 123}+\delta_{ 124}+\delta_{ 1234}}$.
We argue that  ${ \delta_{ 13}<{}\delta_{ 13|2}<{}\delta_{ 13|24}=0}$
because  the information decreases as the conditioning set is enlarged.
Consequently  ${ \delta_{ 123}>0}$, and symmetry shows the other 3rd-order
differences are positive.
Also  ${ 0<\delta_{ 134|2}=\delta_{ 134}+\delta_{ 1234}}$, so that ${ \delta_{ 1234}<0}$.
\par{(e) bayesNetA}:
There are two  independences manifest: 
firstly ${ X_{2}\indep{}X_{4}\mid{}X_{1}}$ translates to 
${ \delta_{ 24|1}=0=\delta_{ 24}+\delta_{ 124}}$; secondly 
${ X_{1}\indep{}X_{3}\mid{}X_{24}}$ translates to 
${ \delta_{ 13|24}=0=\delta_{ 13}+\delta_{ 123}+\delta_{ 134}+\delta_{ 1234}}$.
The 3rd-order difference ${ \delta_{ 124}}$ is  positive and
there is a partial synergy at ${ 3}$ as ${ \delta_{ 234|1}<0}$.
\par{(f) bayesNetB}:
There are two  independences: 
${ X_{2}\indep{}X_{4}}$ implies ${ \delta_{ 24}=0}$; secondly 
${ X_{1}\indep{}X_{3}\mid{}X_{24}}$ is again 
${ \delta_{ 13|24}=0=\delta_{ 13}+\delta_{ 123}+\delta_{ 134}+\delta_{ 1234}}$.
There are two marginal synergies at ${ 1}$ and at ${ 3}$ so ${ \delta_{ 124}<0}$
and ${ \delta_{ 234}<0}$.
\par{(g) bayesNetC}:
There are three marginal independences
between pairs of ${ X_1,X_3,X_4}$ with corresponding 2nd-order
differences being zero; and as these variables are mutually independent
the 3rd-order difference is zero too.
There are three marginal synergies at ${ 2}$,
with ${ \delta_{ 123}<0}$, ${ \delta_{ 124}<0}$ and ${ \delta_{ 234}<0}$.
\end{example}
\par
\begin{example} (GP burn-out):
This example was used by \cite{maassen2001suppressor} to illustrate
suppression in the context of path anaysis.
We use it to illustrate forward differences of the entropy
in four dimensions,
Surprisingly we find that there are no synergies in any of
the three dimensional margins nor  any partial synergy in four 
dimensions, and consequently no colliders.
\par
A two wave study of burnout among ${ 207}$ general practitioners measured
levels of the lack of job satisfaction and of burn-out.
The variables here are denoted by ${ js1,js2,bo1,bo2}$, with the numeral
denoting the wave.
The correlation matrix, reported in supplementary material,
shows all marginal correlations to be positive.
 Forward differences of the entropy  higher than the first are 
\begin{center} {\small\begin{tabular}{r |llll lllll} 
subset&\quad&js1:bo1&js1:js2&bo1:js2&js1:bo2&bo1:bo2&js2:bo2&&\\ 
2nd-order fwd.diff&&-191.6&-98.9&-123.9&-114.5&-356.9&-259.2&&\\ 
subset&&js1:bo1:js2&js1:bo1:bo2&js1:js2:bo2&bo1:js2:bo2&&js1:bo1:js2:bo2\\ 
3,4-orders fwd.diff&&66.96&103.7&71.63&118.5&&-61.96\\ 
\end{tabular}}
\end{center}
\par
The 2nd-order differences (pairwise MIs) are all substantial;  the
3rd-order differences are all positive, so clearly there are no
synergies in any three dimensional marginal.
There are two (approximate) linear relations corresponding
to the 2nd-order  statements
\newline\blank\quad
   ${ js1\indep{}bo2\mid{}\{js2,bo2\}}$:
   ${ \delta_{ js1:bo2}+\delta_{ js1:js2:bo2}+\delta_{ js1:bo1:bo2}+
   \delta_{ js1:bo1:js2:bo2}=-1.13}$mbits; and
\newline\blank\quad
   ${ bo1\indep{js2}|\{js1,bo2\}}$:
   ${ \delta_{ bo1:js2}+\delta_{ js1:bo1:js2}+\delta_{ bo1:js2:bo2} +
   \delta_{ js1:bo1:js2:bo2}=-0.40}$mbits.
\newline
This suggests the 4-cycle with graph 
\begin{center}{\small\begin{picture}(100,80)(0,-10)
           \put(-30,40){\makebox(0,0){}}
           \put(0,0){{\circle{20}}\makebox(0,0){${ js1}$}}
           \put(50,0){{\circle{20}}\makebox(0,0){${ js2}$}}
           \put(50,50){{\circle{20}}\makebox(0,0){${ bo2}$}}
           \put(0,50){{\circle{20}}\makebox(0,0){${ bo1}$}}
           \put(  10,0 ) {\line(1,0){30}}
           \put(  10,50 ) {\line(1,0){30}}
           \put(  0,10 ) {\line(0,1){30}}
           \put(  50,10 ) {\line(0,1){30}}
\end{picture}}\end{center}
Standard model fitting using the R-packages
{\sf pcalg, gRim}, or {\sf ggm}
gives the same independence graph.
\par
The context suggests that synergies might be found at one or both of
the second wave nodes: for js2,
${ \delta_{ js1:js2:bo2|bo1}=\delta_{ js1:js2:bo2}+\delta_{ js1:bo1:js2:bo2}=9.67
}$mbits and for bo2,
\newline
${ \delta_{ bo1:js2:bo2}+\delta_{ js1:bo1:js2:bo2}=56.55}$mbits.
However both are positive indicating that this is not the case, and we
conclude there are no suppression effects manifest in the observed
data.
\end{example}
\par
\subsection{\sffamily Higher dimensions}\label{sect:higher}
\par
\begin{example} (A tree averaging structure):
This artificial tree averaging process provides an example of
computing third order forward differences with respect to a given
graph.
The process starts with a founding generation of independent Gaussian
random variables.
Pairs of these are parents to a single child, giving a new generation
of half the size; and the process repeats until only one successor is
left.
The parent-child relation is specified by the parameter ${ \alpha}$ in
{\begin{eqnarray*}
X_{child} = \alpha(X_{par1}+X_{par2})+\epsilon
\end{eqnarray*}}\blank
where ${ \epsilon}$ are independent standard Normal.
The correlation matrix is determined by the parameter ${ \alpha}$.
With ${ 8}$ founders there are ${ p=15}$ variables; so that in principle there are
${ 455}$ subsets of size ${ 3}$ to examine.
\par
The Bayes network generating the process is displayed in Figure
\ref{fig:treeAv}.
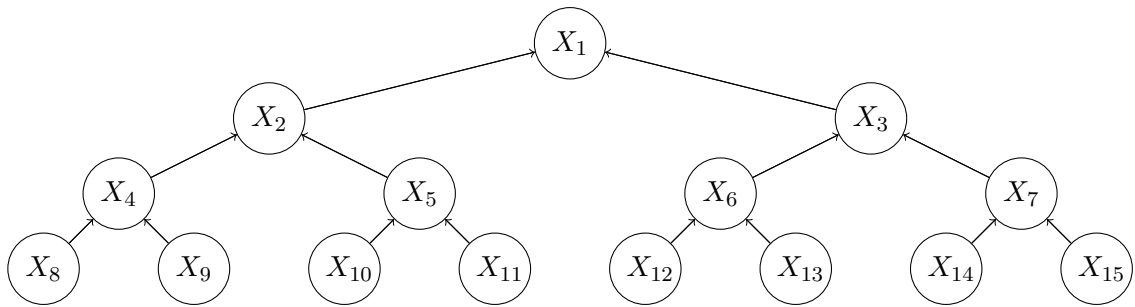
\begin{figure}[!htbp]
\begin{center}{\small
\begin{tikzpicture}
[auto, every node/.style={
  circle,draw, text centered,text width=.5cm,minimum height=.5cm },
  node distance=10cm
];
\tikzset{         level 1/.style={sibling distance = 8cm,level distance=1cm},
         level 2/.style={sibling distance = 4cm,level distance=1cm},
         level 3/.style={sibling distance = 2cm,level distance=1cm}
};
\node (x1) {${ X_{1}}$}
    child {node (x2) {${ X_{2}}$}
            child {node (x4) {${ X_{4}}$ }
                  child {node(x8) {${ X_{8}}$}} child {node (x9){${ X_{9}}$}} }
            child {node (x5) {${ X_{5}}$ } 
                   child {node (x10) {${ X_{10}}$}} child {node(x11) {${ X_{11}}$}} }
           }
    child {node (x3){${ X_{3}}$}
            child {node(x6) {${ X_{6}}$}  
                  child {node(x12) {${ X_{12}}$}} child {node(x13) {${ X_{13}}$}} }
            child {node (x7){${ X_{7}}$}
                  child {node(x14) {${ X_{14}}$}} child {node(x15) {${ X_{15}}$}} }
           }
;
\draw[->,black] (x2)--(x1); \draw[->,black](x3)--(x1);
\draw[->,black] (x4)--(x2); \draw[->,black](x5)--(x2);
\draw[->,black] (x8)--(x4); \draw[->,black](x9)--(x4);
\draw[->,black] (x6)--(x3); \draw[->,black](x7)--(x3);
\draw[->,black] (x10)--(x5); \draw[->,black](x11)--(x5);
\draw[->,black] (x12)--(x6); \draw[->,black](x13)--(x6);
\draw[->,black] (x14)--(x7); \draw[->,black](x15)--(x7);
\end{tikzpicture}
\caption{The Bayes network generating the tree averaging process.
     \label{fig:treeAv}}
}\end{center}
\end{figure}
\par
Emulating a data processing exercise with observations on this process
would lead to the skeleton with ${ 19}$ triples or to the moral graph with
${ 19+12=31}$ triples.
Recall that a triple with respect to a graph, is a subset of size ${ 3}$
with (at least) one node adjacent to the two others.
\par
In the moral graph there are seven synergies (negative third order
forward differences) that exactly correspond to the seven immoralities
in the graph.
With ${ \alpha=0.6}$, the strongest synergy is at the apex of the pyramid
(-164.02mbits), followed by two in next tier (-116.78mbits) and the
four weaker ones at the bottom tier (-53.66mbits).
The positive differences each correspond to a
child${ -}$parent${ -}$grandparent conditional independence.
The four stronger ones (65.90mbits) are at the apex of the tree
and involve the final survivor ${ X_1}$; the other eight positive ones
(44.06mbits) involve a founder node.
\par There are exactly twelve differences that are identically zero
corresponding to moralisation: applying d-separation, for instance to
the 2,3,4 triple, marginally ${ X_{24}\indep{}X_{3}}$, so that both
${ \I_{ 23}}$ and ${ \I_{ 23\mid{}4}}$ are zero.
In large graphs it is more efficient to compute low order forward
differences from the skeleton rather than from the moralised graph of
a Bayes network.
\end{example}
\par
\bNotBmka
\begin{example} (Carcass data):
A well known data set is the so-called carcass data available from the
R-package {\sf gRim}, \cite{hojsgaard2012graphical}; the correlation
matrix is reproduced in supplementary material.
It consists of ${ 7}$ nutritional content measurements on ${ 374}$  pigs(?).
The skeleton is found using the {\sf pcalg} R-package,
\cite{kalisch2012causal}, with standard settings and a ${ 5}$\%
significance level for edge testing, gives the graph of the skeleton
as the left diagram in Figure~\ref{fig:carcass}.
With this graph there are exactly nine node clusters of order ${ 3}$
(triples).
The corresponding forward differences are listed in
Table~\ref{tab:carcass}.
Most of entries are positive, and quite a few are large indicating 
large duplication of effects, especially within the Fat measures
and within the Meat measures.
Strikingly there are two overlapping synergies (the negative 
differences).
They have the same collider LeanMeat, and the nodes of the synergies
are coloured in the graph on the right using the colouring rule above.
\begin{table}[!hbt]
\begin{center}
\caption{Third order forward differences for the carcass data based on the 
graph.
\label{tab:carcass}}
{\scriptsize\begin{tabular} {r lll r}
\\[2pt]
&\multicolumn{3}{c}{Nodes} & ${ \delta}$\\
\hline\\
&Fat11&Meat13&LeanMeat&-78.13\\ 
&Fat12&Meat13&LeanMeat&-76.55\\ 
&Meat12&Meat13&LeanMeat&32.32\\ 
&Meat11&Meat13&LeanMeat&50.54\\ 
&Fat12&Fat13&LeanMeat&458.18\\ 
&Fat11&Fat13&LeanMeat&460.97\\ 
&Fat11&Fat12&LeanMeat&514.10\\ 
&Fat11&Fat12&Fat13&694.67\\ 
&Meat11&Meat12&Meat13&894.42
\end{tabular} }
\end{center}
\end{table}
Reading from the graph Fat11 and Meat13 are marginally independent
and together enhance LeanMeat more than their separate effects would 
warrant. 
The same is true of  the effect of Fat12 and Meat13 on LeanMeat.
Both of these synergies suggest that the data be modelled as a chain
graph, \cite{wermuth1990substantive}, with LeanMeat as the single
outcome variable.
\begin{figure}[!hbt]
\begin{center}
{\includegraphics[width=12cm]{\epsdir/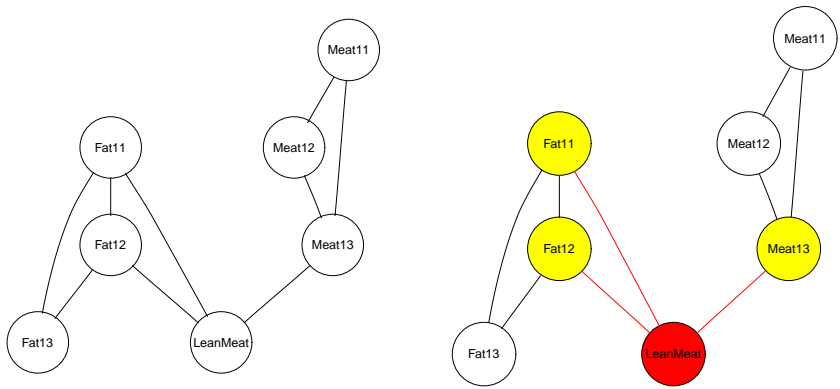}}
\end{center}
\caption{Skeleton of the carcass data (left) with 
two overlapping coloured synergies (right).
\label{fig:carcass}}
\end{figure}
\par
Going from the coloured graph may be misleading without access to the
corresponding table of synergies; for instance the graph might be
taken to indicate that \{Fat11,Fat12,LeanMeat\} is a synergy when it is
not.
\end{example}
\eNotBmka
\par
\begin{example} (Wine quality data):
We consider a  regression example of wine quality taken from the 
machine learning data set repository at UCI
({\sf archive.ics.uci.edu/ml/datasets/ Wine+Quality}).
There are ${ 4898}$ observations on ${ 11}$ physico-chemical properties and a
sensory quality variable for the white Portuguese Vinho Verde wine
reported by \cite{cortez2009modeling}.
The red wine data was used as one of the test sets in
\cite{elidan2010copula}.
The quality outcome is an ordered categorical response, the other
variables are continuous.
Our objective is to find and display any synergies in the explanatory
variables so leading to a better understanding of the data set.
\par
An exploratory analysis reveals transformations are required
to establish linearity and normality.
The simple approach of taking the normal scores, based on ranking each
variable, produces pairs plots for the  bivariate margins that are now
almost all uniformly ovaloid.
\par The skeleton is found using the {\sf pcalg} R-package,
\cite{kalisch2012causal}, with standard settings and a ${ 1}$\%
significance level for edge testing.
There are ${ 21}$ edges and ${ 51}$ triples in the skeleton compared to ${ 55}$
and ${ 165}$, respectively, in the complete graph.
The empirical cumulative distribution function of the corresponding
3rd-order forward differences is displayed on the left in Figure
\ref{fig:wine}.
\begin{figure}[!hbt]
\begin{center}
{\includegraphics[width=12cm]{\epsdir/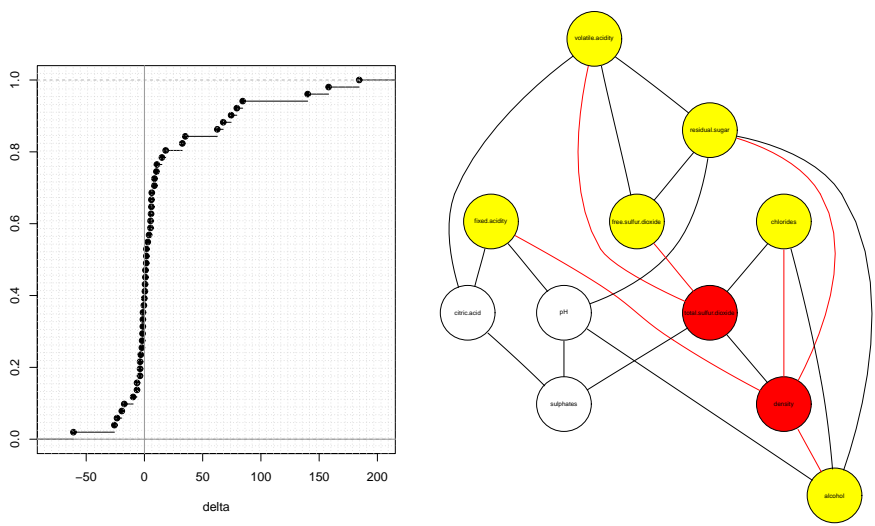}}
\end{center}
\caption{Left: empirical cdf of the 3rd-order differences 
computed from the estimated skeleton of the wine data;
Right: the skeleton of the wine data with  synergies (${ <}$${ -15}$mbits)
coloured.
\label{fig:wine}}
\end{figure}
The majority of the 3rd-order differences are near zero.
There is clearly one large synergy, four others of some size, three large
positive forward differences, and five more of some size.
The detail is given in  Table \ref{tab:wine}.
\begin{table}[!hbt]
\caption{Larger synergies for the wine data, with the node to be coloured
         red indicated by the asterisk.
    \label{tab:wine}}
\begin{center}{\scriptsize\begin{tabular} {rrrr r} 
                \multicolumn{3}{c}{triple} & 3rd-order diff.\\[1pt] 
               \hline\\[-8pt]
residual.sugar&density*&alcohol&-61.74\\ 
volatile.acidity&free.sulfur.dioxide&total.sulfur.dioxide*&-25.49\\ 
fixed.acidity&residual.sugar&density*&-23.29\\ 
fixed.acidity&density*&alcohol&-19.64\\ 
residual.sugar&chlorides&density*&-18.50\\ \\[-8pt]
chlorides&total.sulfur.dioxide&alcohol&79.46\\ 
chlorides&total.sulfur.dioxide&density&84.40\\ 
residual.sugar&total.sulfur.dioxide&density&139.82\\ 
total.sulfur.dioxide&density&alcohol&158.78\\ 
chlorides&density&alcohol&184.43\\[1pt] 
\end{tabular} }\end{center}
\end{table}
\par
The skeleton on the right of the Figure is coloured with the five
synergies in the Table that are stronger than ${ -15}$mbits.
This has the effect of classifying the explanatory variables into red,
yellow and white.
Each red node  belongs to one or more synergistic triples and 
is defined as the node opposite the weakest edge.
There are just two red nodes: density, which occurs in four synergies, and
total.sulfur.dioxide  occurring once.
Three of the four synergies including density are immoralities and
make density an unshielded collider.
The context of this physico-chemical data set suggests a causal
mechanism in which density and total.sulfur.dioxide are responses
directly affected by the yellow nodes in a synergistic relation.
Interestingly  density and total.sulfur.dioxide are associated,
but not  synergistically,  occurring together in the three of the
largest five positive forward differences.
The white nodes are not members of any synergistic triple.
This coloured classification of nodes suggests further fitting
the data as variables in a chain graph, \cite{wermuth1990substantive}.
\end{example}
\par
\section{\sffamily\large Discussion}
\par 
To turn forward difference estimation into a practical statistical
tool requires a  reliable method of assessing sampling errors.
This is clearly necessary in empirical estimation though perhaps less
so in the testing scenarios of graphical model search.
For now we make two remarks.
Firstly, it is probable that lower order conditional mutual
informations have smaller sampling errors than any associated higher
order measures,  as these have fewer additional terms in their
expansion. 
Secondly the sampling error of the highest order order term in the
forward difference expansion of information is probably of the same
order of variability as the information itself.
However in the absence of good approximations to sampling errors the
parametric bootstrap should work well.
\par 
Forward differences of the entropy function promise a productive vein
of research related to graphical models.
For instance the additive expansion of the conditional mutual
information statistic in terms of 3rd-order differences give a
particularly simple proof of the so-called information inequality.
The potential  efficiency gains in graphical model constraint based search 
might be leveraged to attain or surpass that of current algorithms 
such as {\sf pcalg} mentioned above.
A difficult problem is to locate and evaluate higher dimensional
synergies of the form ${ \delta_{ ijk|A}<0}$ where the subset ${ \aa}$ is
abitrary.
A possible line of research is investigation if synergies for shielded
colliders have a role to play in understanding causal graphs.
\par
Parallel to forward differences are backward differences generated by
inverting the lattice of entropies and taking ${ h_{\pp}}$ as the minimal
and ${ h_{\phi}}$ as the maximal elements respectively.
A better way to study this might be to take the forward differences of
the conditional entropy function ${ h_{\pp|{}A}(\pp|A)}$  on
${ \{A\in\calP\}}$.
It is quickly seen that the 2nd-order differences are pair-wise mutual
informations conditioned on the all other variables, and 3rd-order
differences are ${ \delta_{ ijk|\pp\exc{}ijk}}$.
\par
\section*{\sffamily\large Acknowledgement}
We thank Peter van der Heijden for pointing out the connection 
with suppressor variables.  
\bNotBmka
The work of two authors (FM,YX) was supported
by Philip Morris International Research and Development. 
\eNotBmka
\appendix
\par
\section*{\sffamily\large Appendix Proofs}
\begin{proof}{\em\sf Proof of Theorem \ref{thm:symm}.}
The notation ${ \delta_{ \aa}}$ is shorthand for ${ \delta_{ (i;i\in\aa)}}$
where the round brackets indicate an ordered sequence.
We wish to show ${ \delta_{ \pi(\aa)}=\delta_{ \aa}}$ for any permutation ${ \pi}$.
\par
We argue by enumeration on ${ |\aa|}$.
For ${ |\aa|=1}$ there is nothing to show.
For ${ |\aa|=2}$ with ${ \aa=\{i,j\}}$ say
${ h_{\aa}=\delta_{ \phi}+\delta_{ i}+\delta_{ j}+\delta_{ ij}}$ and
${ h_{\pi(\aa)}=\delta_{ \phi}+\delta_{ i}+\delta_{ j}+\delta_{ \pi(ij)}}$.
As the entropies are equal, subtraction shows ${ \delta_{ \pi(ij)}=\delta_{ ij}}$
so that the 2nd-order forward differences are symmetric.
For ${ |\aa|=3}$ a similar result is attained using the symmetry of 
the 2nd-order terms.
The argument continues until ${ |\aa|=p}$.
\end{proof}
\begin{proof}{\em\sf Proof of Lemma \ref{lemm:addit}.}
Note that ${ \delta_{ \phi}=0}$ but the term is included to preserve
symmetry.
The forward difference expansion (\ref{eq:fwddiff}) of ${ h_{a\cup{}b}}$ is
{\begin{eqnarray*}
h_{a\cup{}b} 
= \sum_{c\subseteq{}a\cup{}b} \delta_{ c}
= -\delta_{ \phi}+ h_{a}+ h_{b}+\sum_{c;|a\cap{}c|>1, |b\cap{}c|>1} \delta_{ c}.
\end{eqnarray*}}\blank
The last summation on the right is the sum over terms with at least one element
from ${ a}$ and one from ${ b}$.
By hypothesis it is ${ 0}$.
\par 
Direct enumeration on the elements
${ (|a|,|b|)\in\{1,2,\dots,|A|\}\times\{1,2,\dots,|B|\}}$
shows that every ${ \delta_{ c}}$ in this summation is ${ 0}$.
Start with singletons ${ a=\{i\}}$, ${ b=\{j\}}$.  
The only term is ${ \delta_{ ij}}$ and so it is ${ 0}$.
Repeat this over all pairs ${ ij}$.
A similar argument applied to ${ a=\{i\}}$, ${ b=\{j,k\}}$
and using ${ \delta_{ ij}=0}$
establishes ${ \delta_{ ijk}=0}$, for all ${ k}$.
Repeating  this argument establishes ${ \delta_{ ib}=0}$ 
for any nonempty ${ b\subseteq\bb}$.
A similar enumeration on ${ |a|}$ then gives the result.
\par
The proof of the converse follows immediately from the 
expansion of ${ h_{a\cup{}b}}$.
\end{proof}
\begin{proof}{\em\sf Proof of Theorem \ref{thm:cmi}.}
Take ${ \aa}$ disjoint from ${ \{i,j\}}$ and  note
{\begin{eqnarray}\label{eq:recurr}
h_{\aa{}j} 
&=&  h_{\aa} +  \sum_{j\subseteq{}\bb\subseteq{}\aa{}j} \delta_{ \bb}.
\end{eqnarray}}\blank\newline
This additivity recurrence follows directly from the definition of the
forward differences at (\ref{eq:fwddiff}).
Now use this in the elementary imset representation at (\ref{eq:cmih})
{\begin{eqnarray*}
 -  \I_{ ij|\aa} 
&=&  (h_{\aa{}ij}-h_{\aa{}i})-(h_{\aa{}j}-h_{\aa})  \\
&=&  \sum_{j\subseteq{}\bb\subseteq{}\aa{}ij}        \delta_{ \bb} 
 - \sum_{j\subseteq{}\bb\subseteq{}\aa{}j}          \delta_{ \bb},
\qbox{using (\ref{eq:recurr}),}\\
&=& \sum_{ij\subseteq{}\bb\subseteq{}\aa{}ij}        \delta_{ \bb}.
\end{eqnarray*}}\blank
Cancellation leaves only those terms with both ${ i}$ and ${ j}$ in the
subscript, as required.
\end{proof}
\begin{proof}{\em\sf Proof of Theorem \ref{thm:fwdident}.}
Firstly, we show ${ \delta_{ 12|3}=\delta_{ 123}-\delta_{ 12}}$ as
the structure of the proof is contained in this special case.
Take the definition of a conditional forward difference
{\begin{eqnarray*}
\delta_{ 12|3}
&=&h_{12|3}-h_{1|3}-h_{2|3}+h_{\phi|3},\\
&=&h_{123}-h_{13}-h_{23}+h_{3},
\end{eqnarray*}}\blank
 simplified by applying ${ h_{\aa|\bb}=h_{\aa\cup\bb}-h_{\bb}}$
repeatedly and noting that the four ${ h_{\bb}=h_{3}}$ terms cancel.
The term ${ \delta_{ 123}}$ is a sum over the ${ 2^3}$ elements of the power
set ${ \calP(\{1,2,3\})}$ of the signed function ${ h}$.
Partition this into the sum of those elements that contain
${ 3}$ and those that do not, then
{\begin{eqnarray*}
\delta_{ 123}
&=&\sum_{C\subseteq12} (-1)^{2+1-|\cc3|}h_{C3} +
   \sum_{C\subseteq12} (-1)^{2+1-|\cc|}h_{C},\\
&=&\delta_{ 12|3}-\delta_{ 12}
\end{eqnarray*}}\blank
taking care with the signs, and
as required.
\par
More generally consider (\ref{eq:recursion}); from the definition of 
conditional forward differences
{\begin{eqnarray*}
 \delta_{ \aa{}|\bb{}k} 
&=&   \sum_{\cc\subseteq\aa}(-1)^{|\aa|-|\cc|}h_{\cc|\bb{}k}\\
&=&   \sum_{\cc\subseteq\aa}(-1)^{|\aa|-|\cc|}h_{\cc{}k|\bb{}},
\end{eqnarray*}}\blank
where the ${ h_{k|\bb}}$ terms cancel.
The sum for ${ \delta_{ \aa{}k|\bb{}}}$ is partitioned into the 
sum over the power sets including and excluding ${ k}$:
{\begin{eqnarray*}
 \delta_{ \aa{}k|\bb{}} 
&=&   
  \sum_{\cc\subseteq\aa{}}(-1)^{|\aa{}k|-|\cc{}k|}h_{\cc{}k|\bb{}}
+ \sum_{\cc\subseteq\aa}(-1)^{|\aa{}k|-|\cc|}h_{\cc|\bb{}}\\
&=&  \delta_{ \aa{}|\bb{}k}-\delta_{ \aa{}|\bb{}},
\end{eqnarray*}}\blank
as required.
\end{proof}
\begin{proof}{\em\sf Proof of Theorem \ref{thm:sepfwd}.}
When ${ \cc}$ separates ${ \aa}$ and ${ \bb}$ then 
as a consequence of the Markov properties of the graph
${ X_{\aa}\indep{}X_{\bb}\mid{}X_{\cc}}$;
 consequently in turn
${ h_{\aa\cup\bb|\cc}=h_{\aa|\cc}+h_{\bb|\cc}}$.
By a small generalisation of the the additivity Lemma  \ref{lemm:addit}
to incorporate conditioning, 
the result follows.
\end{proof}
\begin{proof}{\em\sf Proof of Theorem \ref{thm:synergy}.}
Endow the set ${ \aa}$ with a total ordering so that
for ${ i\neq{}j\in\aa}$ either ${ i<j}$ or ${ j<i}$.
Apply the information identity, \cite{cover2006elements}, to get
{\begin{eqnarray}\label{eq:infoident}
\inf(X_{k}\indep{}X_{\aa}) &=&
\sum_{j\in\aa} \inf(X_{k}\indep{}X_{j}\mid{}X_{\{i;\,i<j\}}).
\end{eqnarray}}\blank\newline
 Use (\ref{eq:cmidelta}) of Theorem \ref{thm:cmi} to express the
conditional mutual informations in terms of the forward differences, so
{\begin{eqnarray*}
\inf(X_{k}\indep{}X_{\aa}) 
&=& \sum_{j\in\aa} \left[ 
-\sum_{\bb{}\subseteq{}\{i;\,i<j\}} \delta_{ \bb{}kj} \right].
\end{eqnarray*}}\blank
Isolate the 2nd-order differences from sum and  rearrange
the index of summation gives the result:
{\begin{eqnarray*}
\inf(X_{k}\indep{}X_{\aa}) 
&=& - \sum_{j\in\aa} \delta_{ kj}
-  \sum_{j\in\aa}\sum_{\bb{}\subseteq{}\{i;\,i<j\},\mid{}\bb|>1} \delta_{ \bb{}kj} \\
&=& \sum_{j\in\aa} \inf(X_{j}\indep{}X_{k}) 
-\sum_{\bb\subseteq\aa; \mid{}\bb|>1} \delta_{ \bb{}k}.
\end{eqnarray*}}\blank
\end{proof}
\begin{proof}{\em\sf Proof of Theorem \ref{thm:gauss}.}
The expression (\ref{eq:d123gauss1}) may be derived directly by
evaluating the determinants in the Gaussian entropy.
The second statement follows from (\ref{eq:delta2})
and the fact that ${ I_{ 12|3}=-\log(1-\rho_{12|3}^2)/2}$.
\end{proof}
\par
\bibliography{gmbib/gmiams,gmbib/gm,gmbib/gmsearch,gmbib/gmbayesnet,gmbib/gmsuppress,gmbib/biomark}
\newpage
\bNotBmka
\par
\section*{\sffamily\large Supplementary material to Synergy, suppression and immorality}
Jan 17, 2015
\par
\subsection*{\sffamily A node cluster algorithm for computing  forward differences}\label{sect:cluster}
We are given an undirected (simple) graph on ${ p}$ nodes and wish to
compute  forward differences of order ${ \kappa}$ or less, for that
graph.
For ${ \kappa=3}$ the difference ${ \delta_{ ijk}}$ is evaluated, if in the
graph, the node ${ i}$, say, has two neighbours ${ j}$ and ${ k}$,
so that this triple forms a node cluster.
More generally, a subset (of any order) is viable if there is one node
that is a neighbour to all other nodes.
\par
Examples focus on low order differences so  we adopt a breadth first 
computation.
We resolve orderings by choosing the weakest candidates based on the marginal mutual information ${ \{I_{ ij};\quad{}i,j\in\P\}}$.
This makes sense when adapting the algorithm to discard edges.
\par
{\small\begin{algorithm}[!hbtp]
\caption{A node cluster algorithm} \label{algm:cluster}
\begin{tabbing}
\enspace Increment ${ \kappa}$: \hfill{starting with ${ \kappa=3}$.}
\\ \enspace ${ \bullet}${} LOOP on nodes: \hfill{to pass through whole graph.}
\\\quad Choose node with maximum degree, not yet visited.
\\\quad ${ \bullet}${} LOOP on all tuples (length ${ \kappa}$-${ 1}$) of its neighbours: 
\\\qquad visit weakest tuple first,     \hfill{via sum MIs.}
\\\qquad Check tuple forms a node cluster,
\\\qquad  put subset=(node,tuple), 
\\\qquad if subset is new store.
\\\qquad Evaluate the entropy  of the subset, store.
\\\qquad ${ \bullet}${} LOOP on all sub-subsets  of the subset:
\\\qquad\quad evaluate forward difference,  \hfill{using stored entropies.}
\\\qquad\quad If a relevant sub-subset unvisited,
\\\qquad\qquad compute entropy, store,  
\\ \enspace UNTIL all sub-subsets,  tuples, and  nodes visited.{}
\end{tabbing}
\end{algorithm}}
\newpage
\par
\subsection*{\sffamily Correlation matrices}
GP burn-out data~{\small
\begin{verbatim}
      SAT11  BO11 SAT12  BO12
SAT11 1.000 0.478 0.354 0.379
BO11  0.478 1.000 0.393 0.619
SAT12 0.354 0.393 1.000 0.544
BO12  0.379 0.619 0.544 1.000
\end{verbatim}

}
\par
\par
Wine data~{\small
\begin{verbatim}
  archive.ics.uci.edu/ml/datasets/Wine+Quality 
  dim(wine)   
4898   12
  qnrank=function(x){ 
    n = length(x) 
    qn = qnorm(seq(1:n)/(n+1))
    return(qn[ rank(x ,ties ="random")])
  }
  xqn = apply(wine,2,qnrank)
  data = xqn[,1:11]  
exclude quality as categorical
  noquote(colnames(data))
 [1] fixed.acidity        volatile.acidity     citric.acid         
 [4] residual.sugar       chlorides            free.sulfur.dioxide 
 [7] total.sulfur.dioxide density              pH                  
[10] sulphates            alcohol             
  colnames(data)=NULL
  cor(data)
         [,1]      [,2]     [,3]     [,4]     [,5]      [,6]     [,7]
 [1,]  1.00000 -0.030966  0.31742  0.09673  0.08979 -0.037524  0.10117
 [2,] -0.03097  1.000000 -0.16975  0.10579  0.01394 -0.084837  0.11634
 [3,]  0.31742 -0.169748  1.00000  0.04799  0.04927  0.089613  0.10153
 [4,]  0.09673  0.105791  0.04799  1.00000  0.20868  0.319827  0.41631
 [5,]  0.08979  0.013938  0.04927  0.20868  1.00000  0.162736  0.35118
 [6,] -0.03752 -0.084837  0.08961  0.31983  0.16274  1.000000  0.62336
 [7,]  0.10117  0.116342  0.10153  0.41631  0.35118  0.623356  1.00000
 [8,]  0.29980  0.002518  0.12049  0.74793  0.47804  0.299242  0.53097
 [9,] -0.43610 -0.045856 -0.15785 -0.16324 -0.04935  0.009448  0.01014
[10,] -0.01824 -0.034665  0.07433  0.01986  0.09323  0.068181  0.16226
[11,] -0.12775  0.054468 -0.05880 -0.41334 -0.53105 -0.258037 -0.44020
           [,8]      [,9]    [,10]    [,11]
 [1,]  0.299804 -0.436098 -0.01824 -0.12775
 [2,]  0.002518 -0.045856 -0.03467  0.05447
 [3,]  0.120485 -0.157846  0.07433 -0.05880
 [4,]  0.747925 -0.163236  0.01986 -0.41334
 [5,]  0.478040 -0.049348  0.09323 -0.53105
 [6,]  0.299242  0.009448  0.06818 -0.25804
 [7,]  0.530971  0.010136  0.16226 -0.44020
 [8,]  1.000000 -0.097515  0.11178 -0.80751
 [9,] -0.097515  1.000000  0.15959  0.15053
[10,]  0.111781  0.159591  1.00000 -0.03972
[11,] -0.807506  0.150530 -0.03972  1.00000
\end{verbatim}

}
\par
\eNotBmka

\end{document}